\documentclass[aps,prd,twocolumn,showpacs,superscriptaddress,preprintnumbers,nofootinbib]{revtex4-1}
\usepackage[colorlinks=true, pdfstartview=FitV, linkcolor=red,citecolor=blue, urlcolor=blue,
bookmarks=true, bookmarksnumbered=true, breaklinks]{hyperref}
\usepackage[dvipdfmx]{graphicx}
\usepackage{amsmath,amssymb,tabularx,ulem,mathrsfs,bm}
\begin{document}

\title{Delayed versus accelerated quarkonium formation in a magnetic field}

\author{Kei Suzuki}
\email{k.suzuki.2010@th.phys.titech.ac.jp}
\affiliation{Department of Physics and Institute of Physics and Applied Physics, Yonsei University, Seoul 03722, Korea}

\author{Su Houng Lee}
\email{suhoung@yonsei.ac.kr}
\affiliation{Department of Physics and Institute of Physics and Applied Physics, Yonsei University, Seoul 03722, Korea}

\date{\today}
\preprint{}

\begin{abstract}
Formation time of heavy quarkonia in a homogeneous magnetic field is analyzed by using a phenomenological ansatz of the vector current correlator.
Because the existence of a magnetic field mixes vector quarkonia ($J/\psi$, $\psi^\prime$) and their pseudoscalar partners ($\eta_c$, $\eta_c^\prime$), the properties of the quarkonia can be modified through such a spin mixing.
This means that the formation time of quarkonia is also changed by the magnetic field.
We show the formation time of vector quarkonia is delayed by an idealized constant magnetic field, where the formation time of the excited state becomes longer than that of the ground state.
As a more realistic situation in heavy-ion collisions, effects by a time-dependent magnetic field are also discussed, where delayed formation of $J/\psi$ and $\psi^\prime$ and very early formation of $\eta_c$ and $\eta_c^\prime$ are found.
\end{abstract}
\maketitle

\section{Introduction}

An intense magnetic field is expected to be produced by peripheral heavy-ion collision experiments at the BNL Relativistic Heavy Ion Collider (RHIC) and at the CERN Large Hadron Collider (LHC) \cite{Kharzeev:2007jp,Skokov:2009qp,Voronyuk:2011jd,Deng:2012pc}.
The strength of the fields, however, could drop instantaneously and it may be difficult to extract the initial information of such extreme environments.
One of the promising candidates to observe the short-lived magnetic field is through the magnetically-modified dynamics of heavy (charm or bottom) quarks and heavy (charmed or bottomed) hadrons because heavy quarks can be produced from initial hard nucleon-nucleon collisions {\it in a short time} and the quarkonia created from their quarks can be sequentially observed as a meaningful probe such as dilepton spectra.

Mass spectroscopy of heavy quarkonia modified in a magnetic field is theoretically being revealed from potential models \cite{Alford:2013jva,Bonati:2015dka,Suzuki:2016kcs,Yoshida:2016xgm}, an effective Lagrangian approach \cite{Cho:2014exa,Cho:2014loa,Yoshida:2016xgm} and QCD sum rules \cite{Cho:2014exa,Cho:2014loa}.
These findings could be related to other interesting subjects of quantum chromodynamics (QCD) under magnetic field, such as the anisotropic confinement \cite{Miransky:2002rp,Andreichikov:2012xe,Chernodub:2014uua,Bonati:2014ksa,Rougemont:2014efa,Bonati:2015dka,Simonov:2015yka,Bonati:2016kxj}, heavy quark dynamics \cite{Fukushima:2015wck,Finazzo:2016mhm,Das:2016cwd} and heavy-light meson spectra \cite{Machado:2013rta,Machado:2013yaa,Gubler:2015qok,Yoshida:2016xgm} (see Ref.~\cite{Hattori:2016emy} for a recent review).
In a magnetic field, the longitudinal components of the vector quarkonia ($J/\psi$, $\psi^\prime$,...) can mix with the pseudoscalar partners ($\eta_c$, $\eta_c^\prime$,...) \cite{Alford:2013jva,Cho:2014exa,Cho:2014loa}.
As a result of the mixing, in the dilepton ($e^+e^-$ or $\mu^+ \mu^-$) invariant-mass spectra as measured by experiments, not only usual $J/\psi$- and $\psi^\prime$-like peaks but also ``anomalous" $\eta_c$- and $\eta_c^\prime$-like peaks can appear, so that these can be a qualitative (and quantitative) probe of the existence of early magnetic field.
Therefore, we need to quantitatively determine whether or not such a probe can be realized in more realistic situations.
One of the related phenomena is the modification of the quarkonium formation under magnetic field.

To discuss the quarkonium formation, we summarize related time scales below.
In experiments, the pairs of a heavy quark and a heavy antiquark can be produced by initial nucleon-nucleon collisions.
A typical time scale to create the heavy-quark pairs, which is the so-called ``coherence time", is naively expected to be $\tau_c \sim 1/m_{Q\bar{Q}} < 0.1\, \mathrm{fm}/c$ in the rest frame of the heavy-quark pair.
After such a creation, the heavy-quarks propagate in medium for some time and eventually form hadronic bound states such as quarkonia and heavy-light mesons, where the ``formation time" scale for quarkonia in vacuum has been estimated from some approaches \cite{Karsch:1987zw,Blaizot:1988ec,Kharzeev:1999bh} and it is still under debate (e.g., $\tau_f \sim 0.44$ and $0.91\, \mathrm{fm}/c$ for $J/\psi$ and $\psi^\prime$, respectively \cite{Kharzeev:1999bh}).
In heavy-ion collisions at RHIC, we set $\tau=0$ as the onset of the overlapping between two charged nuclei. 
A magnetic field begins growing up and its strength reaches to the maximum value at $\tau \sim 0.05 \, \mathrm{fm}/c$ which corresponds to the maximal overlapping instant of the two nuclei.
After that, it weakens gradually but its strength with the order of $m_\pi^2$ will survive up to $\tau \sim 0.2 \, \mathrm{fm}/c$ (see, e.g., Refs.~\cite{Voronyuk:2011jd,Deng:2012pc} for details).
Thus, the time scales of quarkonium formation can be influenced by the magnetic field at RHIC.

In Ref.~\cite{Kharzeev:1999bh}, the formation time of vector quarkonia can be connected to the space-time current-current correlator.
From this approach, one can discuss the formation of not only the ground state [$J/\psi$ and $\Upsilon(1S)$] but also excited states [$\psi^\prime$, $\Upsilon(2S)$ and so on]. 
Furthermore, this approach can be applied to quarkonium formation in medium such as finite temperature \cite{Song:2013lov} and time-dependent temperature expected in heavy-ion collisions \cite{Song:2015bja} as well as in vacuum, if we can input the form of the in-medium correlator (or spectral function).
In this paper, by using this approach, we focus on the modification of quarkonium formation time by magnetic field effects with the spin mixing.
Then we will discuss both the formation time of vector quarkonia and pseudoscalar ones newly induced from the vector current correlator.

This paper is organized as follows.
In Sec. \ref{Sec_Formalism}, the theoretical approach to investigate quarkonium formation time is described and it is extended to systems at finite magnetic field.
In Sec. \ref{Sec_Results}, our results are shown and the formation time in a magnetic field is discussed.
Section \ref{Sec_Conclusion} is devoted to our conclusion and outlook.

\section{Formalism} \label{Sec_Formalism}
The approach to evaluate quarkonium formation time from the correlation function was developed in Ref.~\cite{Kharzeev:1999bh} and applied to more realistic situations in Refs.~\cite{Song:2013lov,Song:2015bja}.

\subsection{Quarkonium formation time from correlator}
We start with the space-time correlation function $\Pi_{\mu \nu} (x)$ for the heavy-quark vector current $J_\mu (x) \equiv \bar{Q} \gamma_\mu Q$:
\begin{eqnarray}
\Pi_{\mu \nu} (x) &=& \langle 0 | T [J_\mu (x) J_\nu (0)] | 0 \rangle \nonumber\\
&=&  \int \frac{d^4 q}{(2\pi)^4} e^{-iqx} (q_\mu q_\nu - g_{\mu \nu} q^2) \Pi (q^2).
\end{eqnarray}
For the correlator in the momentum space, $\Pi(q^2)$, we can use the following dispersion relation:
\begin{equation}
\Pi (q^2) = \frac{1}{\pi} \int ds \ \frac{\mathrm{Im} \Pi(s)}{s-q^2}.
\end{equation}
Therefore, after contracting the Lorentz indices, the space-time correlator can be rewritten as \cite{Shuryak:1983jg,Kharzeev:1999bh}
\begin{equation}
\Pi (x) \equiv \Pi_\mu^\mu (x) = \frac{3}{\pi} \int ds s \ \mathrm{Im} \Pi(s) D(s,x^2), \label{dispersion}
\end{equation}
where
\begin{equation}
D(s,\tau=-x^2) = \frac{\sqrt{s}}{4\pi^2\tau} K_1(\sqrt{s} \tau),
\end{equation}
is the relativistic causal propagator of a scalar field in the coordinate space at $x^2 \neq 0$ and, $\tau$ and $K_1$ are the Euclidean proper time and the modified Bessel function, respectively.

In this approach, the whole correlator $\Pi(\tau)$ is phenomenologically decomposed into multiple parts, $\Pi_0(\tau) +\Pi_1(\tau) + \cdots$, and they are corresponding to the $i$th states and the continuum, respectively.
Then the {\it fraction} of the $i$th state at a time $\tau$ is defined by \cite{Kharzeev:1999bh}
\begin{eqnarray}
&& F_0(\tau) \equiv \frac{\Pi_0(\tau)}{\Pi(\tau)}, \ \ \ F_1(\tau) \equiv  \frac{\Pi_1(\tau)}{\Pi(\tau) - \Pi_0(\tau)}, \nonumber\\
&& F_2(\tau) \equiv  \frac{\Pi_2(\tau)}{\Pi(\tau) - \Pi_0(\tau) - \Pi_1(\tau)}, \cdots. \label{def_F}
\end{eqnarray}
We note that $F_i \to 1$ at $\tau \to \infty$ as long as the $i$th state is located below the continuum threshold.
The derivative of the fraction with respect to $\tau$ is called {\it distribution} \cite{Kharzeev:1999bh}
\begin{equation}
P_i(\tau) \equiv \frac{dF_i(\tau)}{d\tau}. \label{def_P}
\end{equation}
Finally the averaged {\it formation time} is defined by the expectation value of the distribution \cite{Kharzeev:1999bh}
\begin{equation}
\langle \tau_\mathrm{form} \rangle_i \equiv \frac{\int d\tau \tau P_i(\tau)}{\int d \tau P_i(\tau)}. \label{def_tau}
\end{equation}
The $\mathrm{Im} \Pi(s)$ in vacuum or in magnetic fields, which are substituted into the dispersion relation (\ref{dispersion}), are assumed to be a spectral function as constructed in the next sections.

\subsection{Spectral ansatz in vacuum}
In vacuum, as the imaginary part of the correlation function, we adopt the following ansatz: Poles + continuum,
\begin{eqnarray}
\mathrm{Im} \Pi(s)  &=& \mathrm{Im} \Pi^{\mathrm{pole}} (s) + \mathrm{Im} \Pi^{\mathrm{cont}} (s) \nonumber\\
&=& \sum_i f_i e_Q^2 \delta(s-m_{V_i}^2) + \frac{e_Q^2}{4\pi} \theta(s - s_\mathrm{th}), \label{Im_vac}
\end{eqnarray}
where $f_i$, $m_{V_i}$, $e_Q$, and $\sqrt{s_\mathrm{th}}$ are the residue and mass of the $i$th hadron resonance, the electric charge of the heavy quarks, and the continuum threshold, respectively.
In this work, we consider only $V_0=J/\psi$ and $V_1=\psi^\prime$ located below the $D\bar{D}$ threshold in vacuum.
Then the corresponding residues in vacuum, $f_0 = 0.545 \, \mathrm{GeV}^2$ and $f_1 = 0.276 \, \mathrm{GeV}^2$, can be determined by the partial decay width to the dilepton: $f_i = 3m_{V_i} \Gamma(V_i \to e^+ e^-)/4 \alpha^2 e_Q^2$, where $\alpha$ is the fine-structure constant.
The continuum threshold is twice the $D $ meson mass, $\sqrt{s_\mathrm{th}} = 2m_D$.
After substituting Eq.~(\ref{Im_vac}) into the dispersion relation (\ref{dispersion}), we obtain  
\begin{equation}
\Pi(\tau) =  \sum_i \frac{3 f_i e_Q^2 m_{V_i}^3}{4\pi^3 \tau} K_1(m_i \tau) + \frac{3e_Q^2}{8\pi^4 \tau^6} \int_{\sqrt{s_\mathrm{th}} \tau}^\infty dx x^4 K_1(x). \label{Pi_tau}
\end{equation}

\subsection{Spectral ansatz in a magnetic field} 
Next we construct a spectral ansatz which reproduces the vector spectral function in a magnetic field.
In a magnetic field, we add the induced pseudoscalar ($\eta_c$) poles into vector ($J/\psi$) spectral function.
The imaginary part of the correlator for the vector channel in a magnetic field is assumed to be
\begin{eqnarray}
\mathrm{Im} \Pi^{eB} (s) &=& \sum_{i=0,1} f_i e_Q^2 \left[ \sin^2 \theta_{i, eB} \delta(s-m_{P_i,eB}^2 ) \right. \label{Im_eB} \\
&& \left. + \cos^2 \theta_{i, eB} \delta(s-m_{V_i,eB}^2) \right ] +\mathrm{Im} \Pi^{\mathrm{cont}} (s).  \nonumber
\end{eqnarray}
In this work, we consider only $P_0 =\eta_c$, $P_1=\eta_c^\prime$, $V_0=J/\psi$, and $V_1=\psi^\prime$.
Magnetic field dependences of the poles are introduced as meson masses (or pole shifts) and mixing angles $\theta_{eB}$ which modify the pole residues.
Here, to simplify, we neglect the magnetic field dependence of the continuum. 
We note that, if $eB \to 0$, this functional form agrees with that in vacuum, Eq.~(\ref{Im_vac}).

To obtain charmonium masses in a magnetic field, we use the following matrix form of the equations of motion derived from an effective Lagrangian \cite{Cho:2014exa,Cho:2014loa}.
For $i$th pseudoscalar $P_i$ (longitudinal vector ${V_i}_{\|}$) state with mass $m_{P_i}$ ($m_{V_i}$) in vacuum,
\begin{equation}
\left(
\begin{array}{cc}
- \omega^{2} +  m_{P_i}^{2}  & - i \frac{g_{P_i V_i}}{m_i} \omega eB \\
i \frac{g_{P_i V_i}}{m_{i}} \omega eB & - \omega^{2} +  m_{V_i}^2 \\
\end{array}
\right)
\left(
\begin{array}{c}
P_i \\
{V_i}_{\|} \\
\end{array}
\right)
=0, \label{matrix_formula}
\end{equation}
where $g_{P_i V_i}$ and $m_i = (m_{P_i} + m_{V_i}) /2$ are the dimensionless coupling constant and the averaged mass in vacuum, respectively.
If the determinant of this matrix is zero, except for $\omega=0$, we can reach a mass formula in a magnetic field \cite{Cho:2014exa,Cho:2014loa}:
\begin{equation}
m^2_{i,eB}= \frac{1}{2} \left( M_{i+}^2 +\frac{\gamma^2}{m_i^2} \pm \sqrt{M_{i-}^4  + \frac{2\gamma_i^2 M_{i+}^2}{m_i^2} + \frac{\gamma_i^4}{m_i^4} } \right), \label{mass_formula}
\end{equation}
where $M_{i+}^2 = m_{V_i}^2 + m_{P_i}^2$, $\gamma_i=g_{P_i V_i} eB$ and $M_{i-}^2 =m_{V_i}^2-m_{P_i}^2$.
The different signs ``$\pm$" in Eq.~(\ref{mass_formula}) correspond to the vector and pseudoscalar channels, respectively.
The dimensionless coupling constants, $g_{P_i V_i}$, are estimated by the experimental values of the radiative decay widths of charmonia (see Refs.~\cite{Cho:2014loa,Yoshida:2016xgm} for the detailed procedure): $g_{\gamma \eta_c J/\psi} = 2.0877$ and $g_{\gamma \eta_c^\prime \psi^\prime} = 3.3762$.
The mixing between $1S$ and $2S$ states such as $\eta_c$-$\psi^\prime$ and $\eta_c^\prime$-$J/\psi$ can be neglected as long as the magnetic field is small enough \cite{Yoshida:2016xgm}.

\begin{figure}[t!]
    \centering
    \includegraphics[clip, width=1.0\columnwidth]{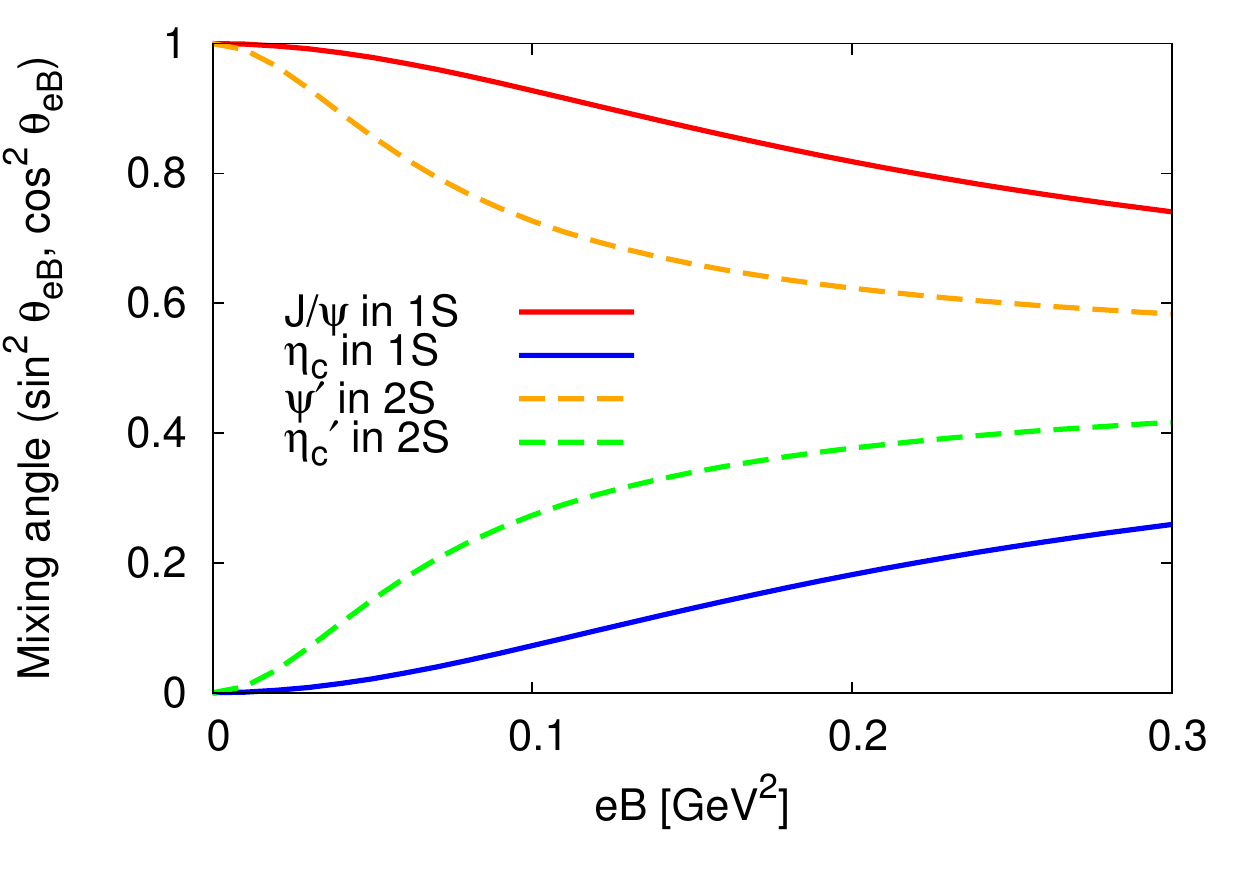}
    \caption{Mixing angles of charmonia in a magnetic field, which are defined by Eqs.~(\ref{angle_def1}) and (\ref{angle_def2}).}
    \label{angle}
\end{figure}

To obtain the mixing angle between the wave functions of pseudoscalar and vector, we define the following form.
After substituting the averaged mass in a magnetic field, $\omega_i = (m_{P_i,eB} + m_{V_i,eB})/2$ into Eq.~(\ref{matrix_formula}), we obtain two eigenvalues and the corresponding eigenvectors, $(iA,B)$ and $(iC,D)$, where $A$, $B$, $C$, and $D$ correspond to the wave functions of the vector or pseudoscalar component in the mixed state, respectively.
Then an approximated mixing angle is defined as follows:
\begin{eqnarray}
&& \sin^2 \theta_{i, eB} \equiv \frac{B^2}{A^2+B^2} = \frac{C^2}{C^2+D^2}, \label{angle_def1} \\
&& \cos^2 \theta_{i, eB} \equiv \frac{A^2}{A^2+B^2} = \frac{D^2}{C^2+D^2}, \label{angle_def2}
\end{eqnarray}
which satisfies the normalization condition $\sin^2 \theta_{i, eB} + \cos^2 \theta_{i, eB} = 1$ for the $i$th pseudoscalar-vector state.
The estimated mixing angle is plotted in Fig.~\ref{angle}.
From these results, we see that the mixing between the excited states is more sensitive than that of the ground states, which means that the residues of poles for the excited states can be also modified more drastically.

After substituting the imaginary part (\ref{Im_eB}) to the dispersion relation (\ref{dispersion}), we finally obtain
\begin{eqnarray}
\Pi^{eB} (\tau) = \sum_{i=0,1} \frac{3 f_i e_Q^2}{4\pi^3} \left[ \frac{ \sin^2 \theta_{i, eB} m_{P_i}^3}{ \tau} K_1( m_{P_i} \tau) \right. \nonumber\\
\left. + \frac{ \cos^2 \theta_{i, eB} m_{V_i}^3}{  \tau} K_1(m_{V_i} \tau) \right] + \Pi^{\mathrm{cont}}(\tau),  \label{Pi_tau_eB}
\end{eqnarray}
where the continuum part $\Pi^{\mathrm{cont}}(\tau)$ is same as that in Eq.~(\ref{Pi_tau}).

\section{Numerical results} \label{Sec_Results}
From the constructed correlator, Eq. (\ref{Pi_tau}) or (\ref{Pi_tau_eB}), we can investigate the fraction $F_i$, distribution $P_i$, and formation time $\langle \tau_\mathrm{form} \rangle_i$ for the $i$th state.
In a finite magnetic field, the mass (or peak position) hierarchy on the spectral function is $m_{\eta_c} < m_{J/\psi} < m_{\eta_c^\prime} < m_{\psi^\prime}$ as long as there is no level crossing (see Refs.~\cite{Suzuki:2016kcs,Yoshida:2016xgm} for a detail).
Therefore, we can define $F_i$ by that order from the original definition (\ref{def_F}).

\subsection{Constant magnetic field}
First we discuss quarkonium formation time in a constant magnetic field.
In vacuum, the lowest state is $J/\psi$ and its contribution dominates the whole correlator at $\tau \to \infty$.
On the other hand, at finite magnetic field, $\eta_c$ becomes the lowest states while $J/\psi$ should behave as an excited state.

\begin{figure}[t!]
    \centering
    \includegraphics[clip, width=1.0\columnwidth]{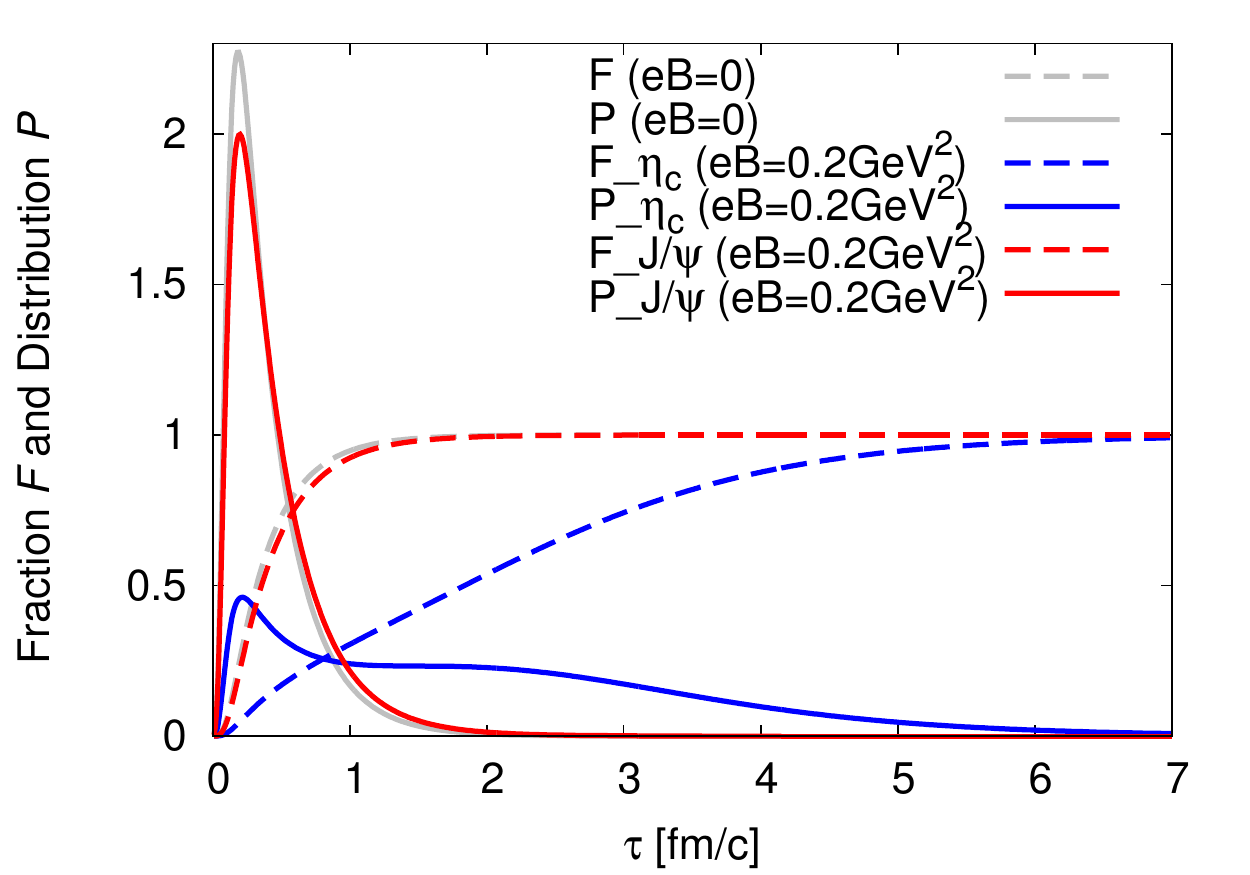}
    \includegraphics[clip, width=1.0\columnwidth]{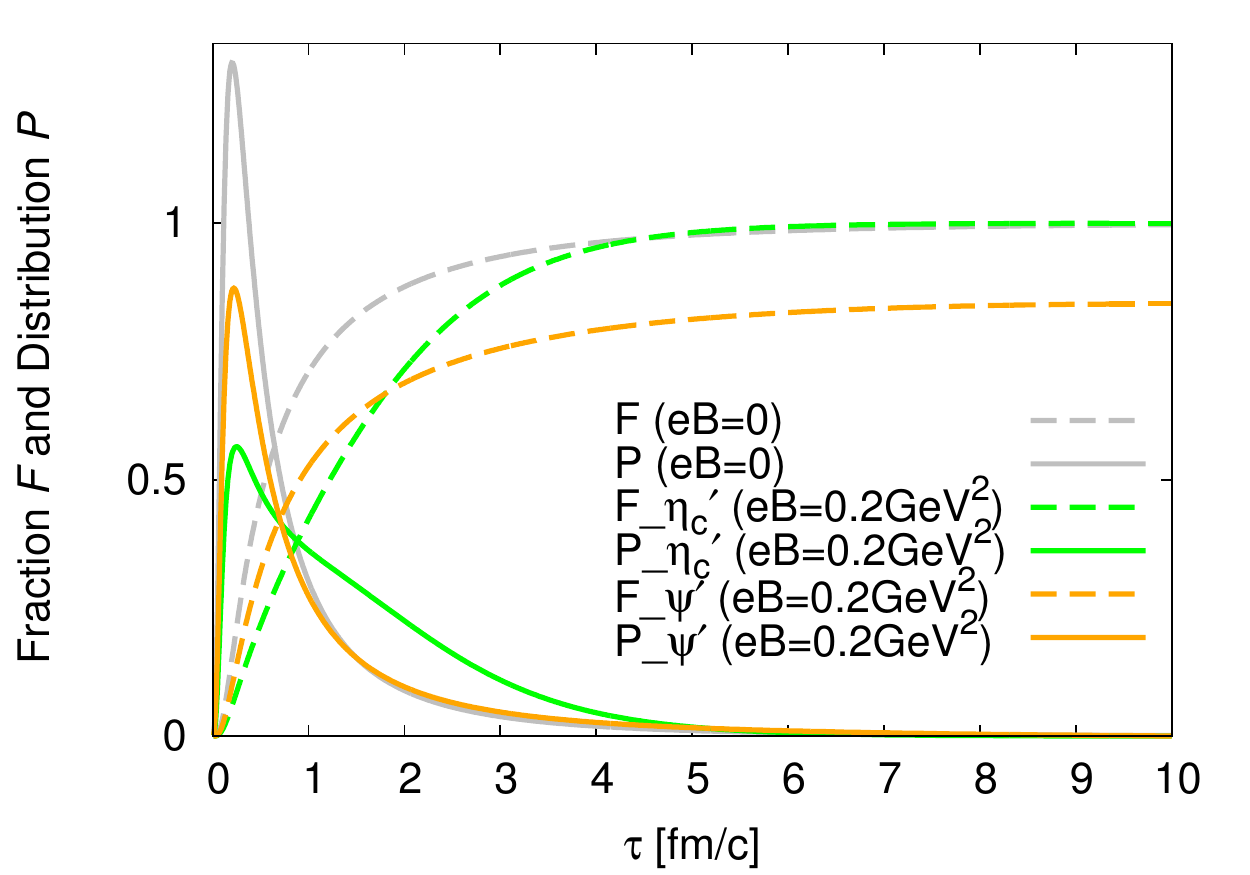}
    \caption{Examples of fraction $F_i$ and distribution $P_i$ for charmonia in a constant magnetic field.
Upper: $J/\psi$ in vacuum and $J/\psi$-$\eta_c$ at $eB=0.2 \, \mathrm{GeV}^2$.
Lower: $\psi^\prime$ in vacuum and $\psi^\prime$-$\eta_c^\prime$ at $eB=0.2 \, \mathrm{GeV}^2$.}
    \label{FandP}
\end{figure}

As an example, the fraction and distribution at $eB=0.2 \, \mathrm{GeV}^2$ are shown in Fig.~\ref{FandP}.
From the upper panel of Fig.~\ref{FandP}, we find that the $F_i$ and $P_i$ for $\eta_c$, as shown by the blue lines, are distributed in the region of $0<\tau<6 \, \mathrm{fm}/c$.
These behaviors are different from those for $J/\psi$ which is located in the faster region of $0<\tau<2 \, \mathrm{fm}/c$.
As a result, the induced $\eta_c$ state has a longer formation time than $J/\psi$ although $\eta_c$ appears as the lowest state.
Such a behavior arises from its small residue in the magnetic field and we expect that, in a larger magnetic field where the mixing becomes strong enough, the formation time of $\eta_c$ becomes shorter.
For $J/\psi$ as shown by the red lines, we find that $F_i$ and $P_i$ is slightly shifted to the larger $\tau$ region, which leads to a longer formation time of $J/\psi$. 

Next we discuss the excited states.
In vacuum, $P_i$ for $\psi^\prime$ is distributed in the wider range of $\tau$ than $J/\psi$, which leads to longer formation time.
In a magnetic field, since the longitudinal component of $\psi^\prime$ mixes with $\eta_c^\prime$, the correlator (after subtracting $\eta_c$ and $J/\psi$) is dominated by both the $\eta_c^\prime$ and $\psi^\prime$.
We comment that $F_i$ for $\eta_c^\prime$ approaches to $F_i \to 1$ at $\tau \to \infty$ while that for $\psi^\prime$ {\it does not}.
This behavior means that the mass of $\psi^\prime$ at $eB=0.2 \, \mathrm{GeV}^2$ exceeds the $D\bar{D}$ threshold.
Then the correlator at $\tau \to \infty$ is dominated not by the $\psi^\prime$ contribution but by the continuum.
Therefore, in this region, we cannot define the formation time of $\psi^\prime$ from our approach, where $\psi^\prime$ is no longer a bound state.

\begin{figure}[t!]
    \centering
    \includegraphics[clip, width=1.0\columnwidth]{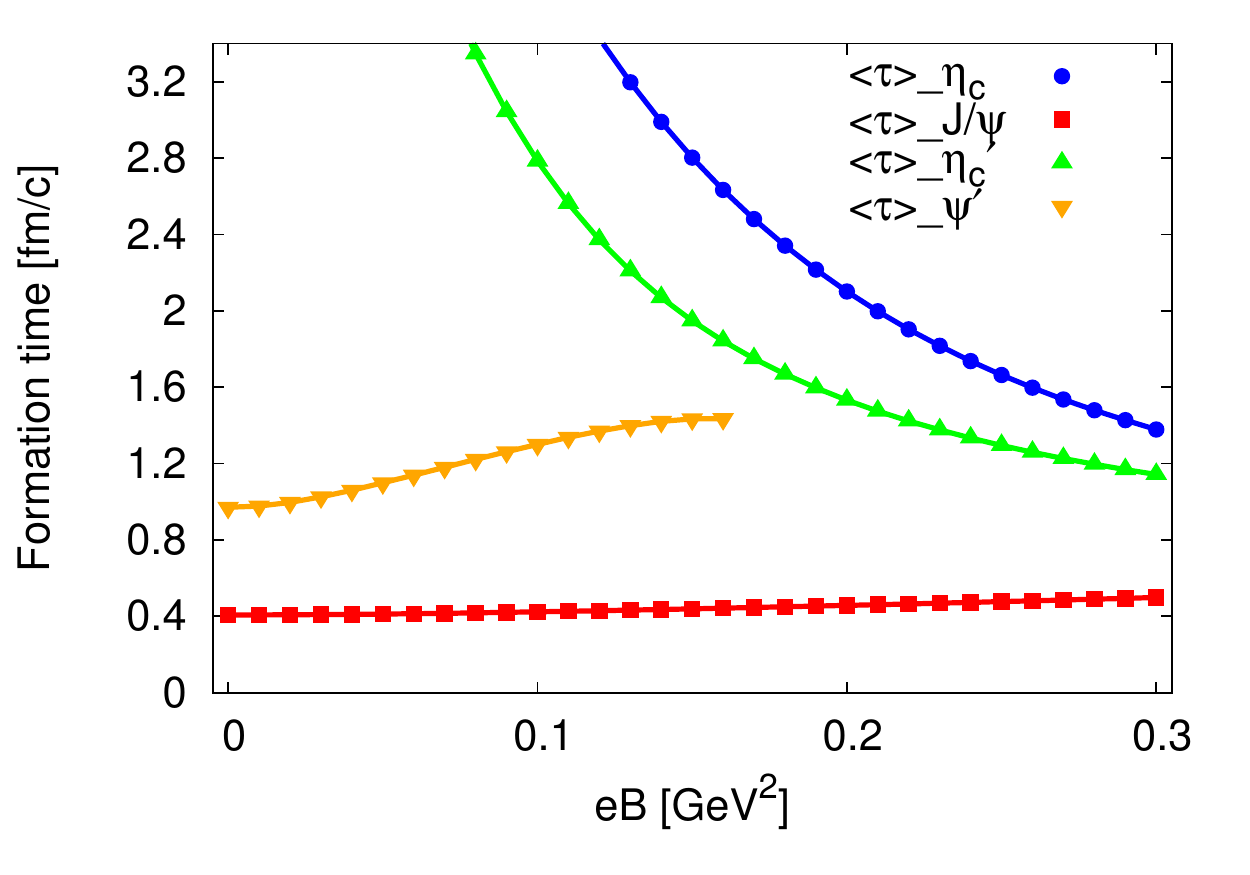}
    \caption{Quarkonium formation time $\langle \tau_\mathrm{form} \rangle_i$ in a constant magnetic field.}
    \label{time}
\end{figure}

The average formation time in a magnetic field is summarized in Fig.~\ref{time}.
We find that the formation times of $J/\psi$ and $\psi^\prime$ become slower with increasing magnetic field.
In particular, the formation time of the excited state $\psi^\prime$ is more sensitive than the ground state.
In contrast to the vector states, the formation time of the pseudoscalar states becomes faster as the mixing in a magnetic field increases their residues.

\subsection{Time-dependent magnetic field}
Next we investigate the effect by the time-dependent magnetic fields as created in relativistic heavy-ion collisions.
We input the time evolution of the magnetic field estimated from the HIJING model in Ref.~\cite{Deng:2012pc}, which corresponds to Au + Au collisions at $\sqrt{s} =200 \, \mathrm{GeV}$ in RHIC.
The maximal magnetic field perpendicular to the reaction plane was estimated to be $eB_{\mathrm{max}} \sim 5 m_\pi^2 \sim 0.1 \, \mathrm{GeV}^2$ for peripheral collisions with the impact parameter $b=10 \, \mathrm{fm}$.
The magnetic field at a time $\tau^\prime$ is parametrized as follows \cite{Huang:2015oca}~\footnote{
Here, $\tau^\prime$ is not Euclidean time but real one.
Since $F_i(\tau)$ defined by Eq.~(\ref{def_F}) is a function of Euclidean time $\tau$, $\tau^\prime$ has to be transformed to $\tau$.
In this work, Eqs.~(\ref{def_PnorP})--(\ref{def_FnorV}) should be regarded as simplified quantities estimated from the magnetic field $eB (\tau^\prime)$ as the snapshot at a real time $\tau^\prime$.
}:
\begin{equation}
eB (\tau^\prime) = \frac{eB_{\mathrm{max}}}{{ \left[ 1+(\tau^\prime - \tau_0)^2/\tau_B^2 \right] }^{3/2}},  \label{eB_tau}
\end{equation}
where $\tau_0=0.05 \, \mathrm{fm}/c$ and $\tau_B=0.065 \, \mathrm{fm}/c$ \cite{Huang:2015oca} are the time from the initial contact to the maximal overlapping between the two nuclei and the lifetime of the magnetic fields, respectively.

Here we define the {\it normalized distribution} and {\it normalized fraction} in a time-dependent magnetic field as follows:
\begin{eqnarray}
&& \left. P^{\mathrm{nor}}_{P_i}(\tau) \equiv \frac{dF_{P_i}(\tau,eB(\tau^\prime)) \sin^2 \theta_{i, eB(\tau^\prime)} }{d\tau} \right|_{\tau^\prime = \tau}, \label{def_PnorP} \\
&& \left. P^{\mathrm{nor}}_{V_i}(\tau) \equiv \frac{dF_{V_i}(\tau,eB(\tau^\prime)) \cos^2 \theta_{i, eB(\tau^\prime)} }{d\tau} \right|_{\tau^\prime = \tau}, \label{def_PnorV} \\
&& F^{\mathrm{nor}}_{P_i}(\tau) \equiv \int_0^\tau d \tau^\prime P^{\mathrm{nor}}_{P_i}(\tau^\prime), \label{def_FnorP} \\
&& F^{\mathrm{nor}}_{V_i}(\tau) \equiv \int_0^\tau d \tau^\prime P^{\mathrm{nor}}_{V_i}(\tau^\prime). \label{def_FnorV}
\end{eqnarray}
The normalization condition is given by $F^{\mathrm{nor}}_{P_i}(\tau) + F^{\mathrm{nor}}_{V_i}(\tau) \simeq 1$ at $\tau \to \infty$.
We note that these definitions are different from those by Eq.~(\ref{def_F}) in which $F_{P_i}(\tau) = 1$ and $F_{V_i}(\tau) = 1$ at $\tau \to \infty$.

\begin{figure}[t!]
    \centering
    \includegraphics[clip, width=1.08\columnwidth]{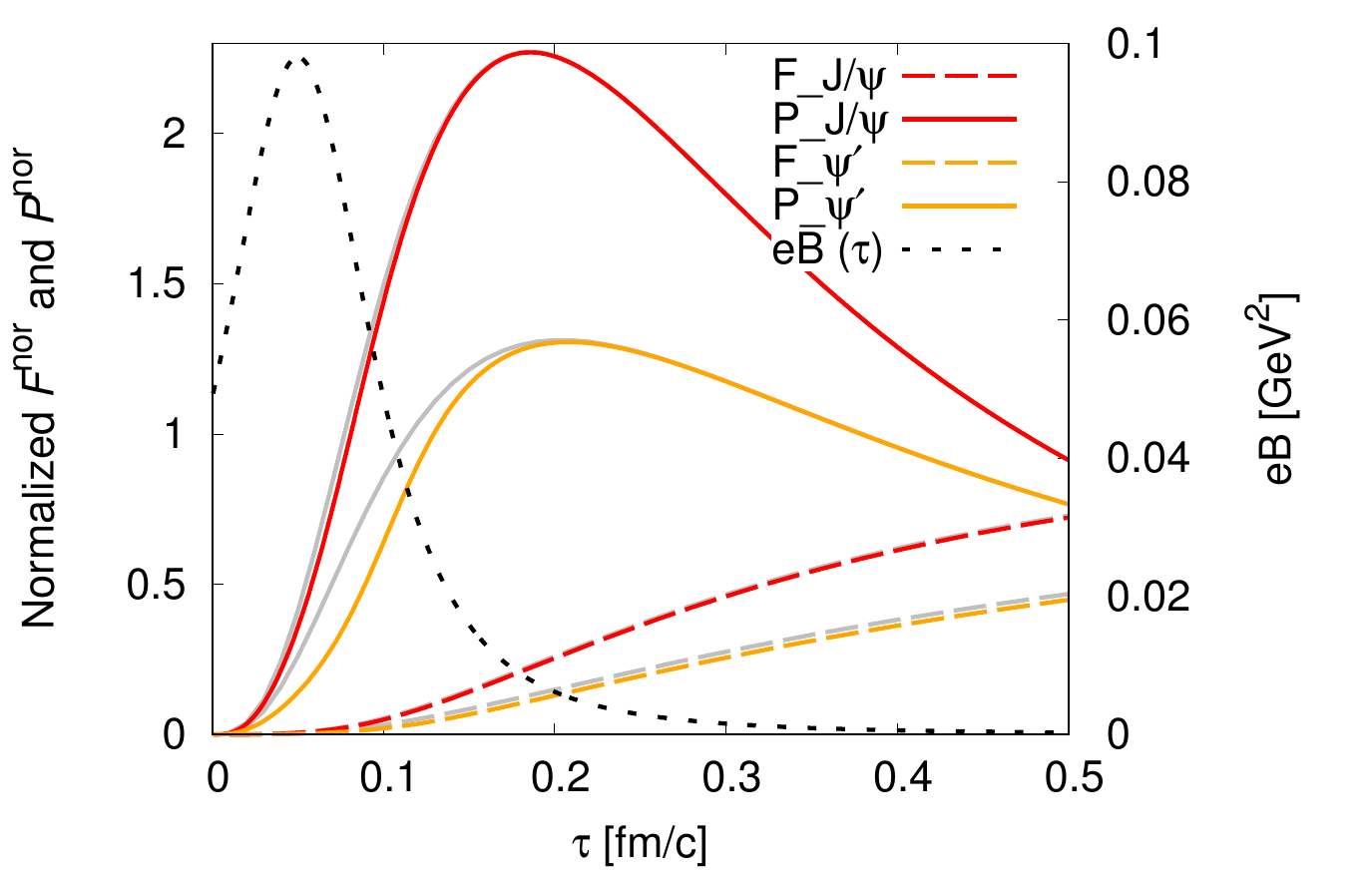}
    \includegraphics[clip, width=1.08\columnwidth]{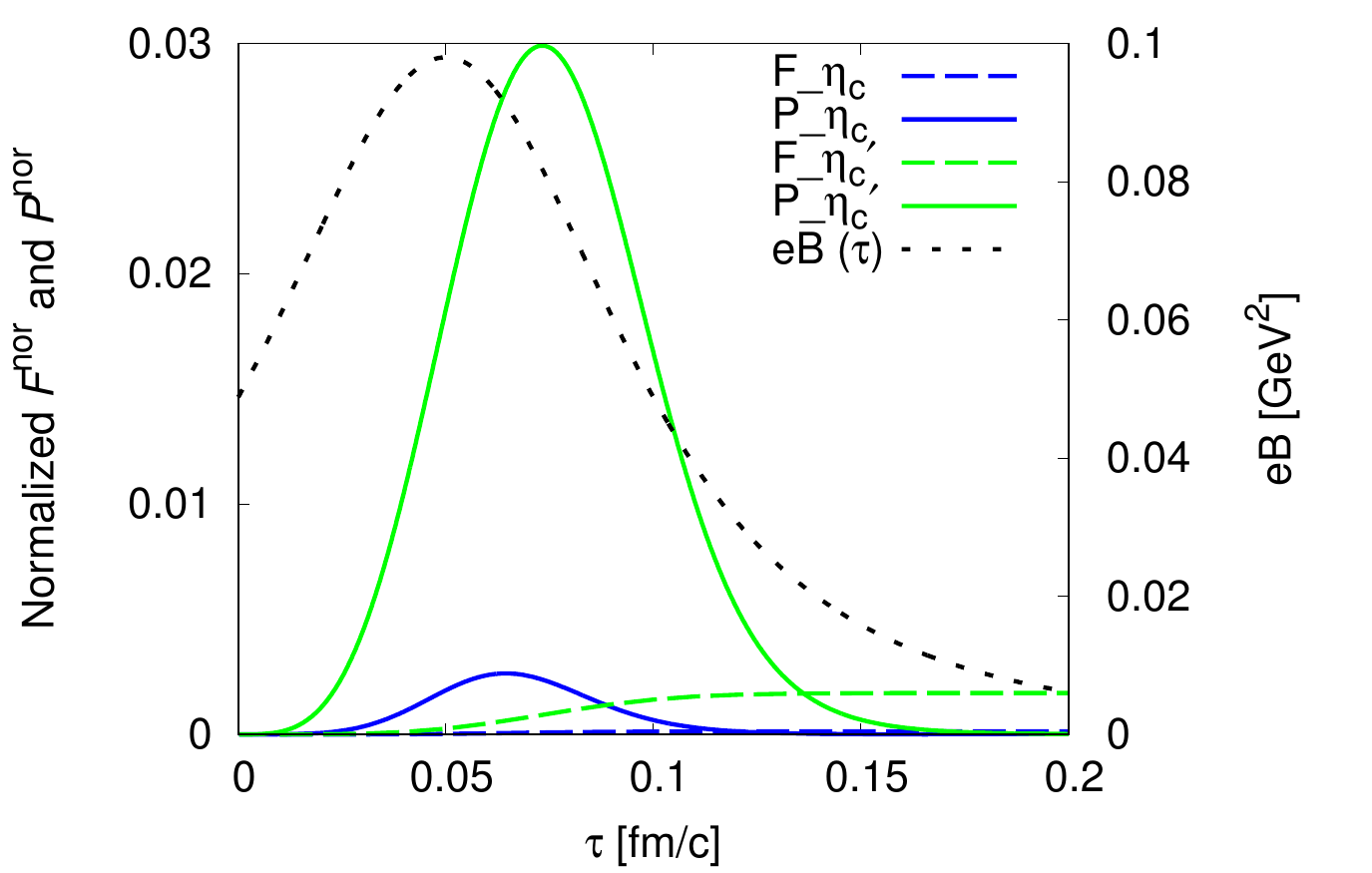}
    \caption{Normalized fraction $F^{\mathrm{nor}}_i$ and distribution $P^{\mathrm{nor}}_i$, defined by Eqs.~(\ref{def_PnorP})--(\ref{def_FnorV}), for charmonia in a time-dependent magnetic field at RHIC.
The black dotted line is the time dependence of magnetic field, given by Eq.~(\ref{eB_tau}).
Upper: $J/\psi$ and $\psi^\prime$. The gray dashed (solid) line stands for $F_i$ ($P_i$) in vacuum.
Lower: $\eta_c$ and $\eta_c^\prime$.}
    \label{norFandP}
\end{figure}

Numerical results are shown in Fig.~\ref{norFandP}.
In the upper figure, we find that the fraction and distribution for $J/\psi$ and $\psi^\prime$ are modified by the time-dependent magnetic field which has the peak of the strength at $\tau \sim 0.05 \, \mathrm{fm}/c$ and is distributed at the time region of $\tau < 0.4 \, \mathrm{fm}/c$.
In particular, the distribution of $\psi^\prime$ is shifted to the later time and the change is larger than that of $J/\psi$, which is caused by the different mixing ratios between the ground and excited states.

In the lower figure, we can see the behaviors of $\eta_c$ and $\eta_c^\prime$ produced from the mixing by the short pulse of the magnetic field. 
The ground state $\eta_c$ is produced at about $\tau \sim 0.065 \, \mathrm{fm}/c$ while the excited state $\eta_c^\prime$ is located at the later time $\tau \sim 0.073 \, \mathrm{fm}/c$.
This is because the distributions (or formation times) of $J/\psi$ which is the mixing partner with $\eta_c$, is faster than those of $\psi^\prime$.
Furthermore, the strength of the distribution $\eta_c^\prime$ is larger than that of $\eta_c$ because of their mixing rates.
At $\tau \to \infty$, the normalized fractions, $F^{\mathrm{nor}}_{P_i}(\tau)$, of $\eta_c$ and $\eta_c^\prime$ reach at saturation values of $0.01\%$ and $0.2\%$, respectively.
These values correspond to the ratios of the pseudoscalar charmonia created only from the vector current correlator: $0.01\%$ of the original $J/\psi$ can be transformed into $\eta_c$ and $0.2\%$ of $\psi^\prime$ into $\eta_c^\prime$.

It should be noted that the yield of $\eta_c$ and $\eta_c^\prime$ estimated in this work is a part of all the $\eta_c$ production.
Actually, almost all of the $\eta_c$ can be also produced from the usual pseudoscalar current correlator.
At the same time, we can also expect a few $J/\psi$ production through the mixing in the pseudoscalar correlator.

In particular, magnetically-induced $\eta_c$ and $\eta_c^\prime$ components from the vector current can show ``anomalous" decay modes such as $\eta_c, \eta_c^\prime \rightarrow e^+ e^- (\mu^+ \mu^-)$ in the presence of an external magnetic field.
Therefore, in heavy-ion collisions, we can observe early formation of $\eta_c$ and $\eta_c^\prime$ from the vector current in the dilepton spectra.
On the other hand, the $J/\psi$ and $\psi^\prime$ components formed from the pseudoscalar current through mixing can decay to dileptons.
Thus, we can identify the difference between charmonia created from the vector or pseudoscalar currents.

As a result, the quarkonium formation times in the magnetic field are estimated as $\langle \tau_\mathrm{form} \rangle_{eB} = 0.086$, $0.408$, $0.095$ and $0.980 \, \mathrm{fm}/c$ for $\eta_c$, $J/\psi$, $\eta_c^\prime$ and $\psi^\prime$, respectively.
Here we find that the formation times of $J/\psi$ and $\psi^\prime$ become slightly slower than the values in vacuum, $\langle \tau_\mathrm{form} \rangle_{J/\psi} = 0.407 \, \mathrm{fm}/c$ and $\langle \tau_\mathrm{form} \rangle_{\psi^\prime} = 0.971 \, \mathrm{fm}/c$.
In addition, we emphasize that $\eta_c$ and $\eta_c^\prime$ from the vector current can appear at very early time because of the existence of the very early magnetic field.
Therefore, the early formed mixed states (or $\eta_c$ and $\eta_c^\prime$) might carry information about early stage of the collision before the thermalization.

We note that our analysis does not take into account the electric conductivity of the thermal medium (or quark-gluon plasma) created after the collisions.
The electric conductivity can increase the lifetime of a magnetic field \cite{Tuchin:2013ie,McLerran:2013hla,Tuchin:2013apa,Gursoy:2014aka,Zakharov:2014dia,Tuchin:2015oka}, so that the quarkonium formation time in more realistic situations could be longer.
In particular, it could be important to adopt the magnetic field (or time) dependence of the conductivity \cite{Buividovich:2010tn,Hattori:2016cnt,Hattori:2016lqx}.

\section{Conclusion and outlook} \label{Sec_Conclusion}
In this paper, we investigated the quarkonium formation time modified in a magnetic field from the space-time correlator of the heavy-quark vector current.
Under constant magnetic field, we found that the formation times of the vector charmonia become slower, while those of the pseudoscalar partners become faster with increasing magnetic field.
In the time-dependent magnetic field which is more realistic at RHIC, we showed the slightly slow formation time for vector charmonia and the early formation of pseudoscalar charmonia from the vector current correlator.
Such early formed $\eta_c$ and $\eta_c^\prime$ can be promising observables in the dilepton spectra as a probe of the existence of early magnetic field.

Furthermore, we have established the new formalism with ``a normalized distribution" and ``a normalized fraction" as Eqs.~(\ref{def_PnorP})--(\ref{def_FnorV}).
It enables us to quantitatively estimate the absolute fraction of the vector and pseudoscalar quarkonia produced from the initially created heavy-quark pair.
Another important point is that although we have estimated the production and mixing from the evolution of the vector current, it is essentially calculating the evolution and formation from a color-singlet charm-quark pair produced at the same point with spin 1.
Hence, our result represents a typical example of how the effect of magnetic fields will influence the formation time from any initial production of heavy-quark pair, so that it will be useful for investigations of the other channels.

It is also interesting to investigate the interplay between hot medium (or QGP) and magnetic field effects
in relativistic heavy-ion collisions. For example, if $J/\psi$ formation time becomes longer by a magnetic field,
heavy quarks with a low momentum can pass through the medium before the thermalization, and the formed
$J/\psi$ could not been affected by the initial hot thermal effects, which could lead to the enhancement of $J/\psi$ survivability from
temperature effects. To discuss such thermal effects for formation time, we can make use of a hybrid approach
between this work and Refs.~\cite{Song:2013lov,Song:2015bja}, which will be a topic for future studies.

Our results can provide new ingredients to the $J/\psi$ production \cite{Yang:2011cz,Machado:2013rta,Guo:2015nsa} and collective flow of heavy flavors \cite{Guo:2015nsa,Fukushima:2015wck,Das:2016cwd} under magnetic field.
It may be also important to discuss how our results influence $J/\psi$ suppression at finite temperature and magnetic field, as discussed in Refs.~\cite{Dudal:2014jfa,Rougemont:2014efa,Bonati:2016kxj}, and the dissociation from other processes such as the Lorentz ionization \cite{Marasinghe:2011bt} and the anomalous flow by the chiral anomaly \cite{Sadofyev:2015hxa}.

\begin{acknowledgments}
The work was supported by the Korea National Research Foundation under Grant No. KRF-2011-0030621 and the Korean Ministry of Education under Grant No. 2016R1D1A1B03930089.
\end{acknowledgments}

\bibliography{FT}

\begin{thebibliography}{45}%
\makeatletter
\providecommand \@ifxundefined [1]{%
 \@ifx{#1\undefined}
}%
\providecommand \@ifnum [1]{%
 \ifnum #1\expandafter \@firstoftwo
 \else \expandafter \@secondoftwo
 \fi
}%
\providecommand \@ifx [1]{%
 \ifx #1\expandafter \@firstoftwo
 \else \expandafter \@secondoftwo
 \fi
}%
\providecommand \natexlab [1]{#1}%
\providecommand \enquote  [1]{``#1''}%
\providecommand \bibnamefont  [1]{#1}%
\providecommand \bibfnamefont [1]{#1}%
\providecommand \citenamefont [1]{#1}%
\providecommand \href@noop [0]{\@secondoftwo}%
\providecommand \href [0]{\begingroup \@sanitize@url \@href}%
\providecommand \@href[1]{\@@startlink{#1}\@@href}%
\providecommand \@@href[1]{\endgroup#1\@@endlink}%
\providecommand \@sanitize@url [0]{\catcode `\\12\catcode `\$12\catcode
  `\&12\catcode `\#12\catcode `\^12\catcode `\_12\catcode `\%12\relax}%
\providecommand \@@startlink[1]{}%
\providecommand \@@endlink[0]{}%
\providecommand \url  [0]{\begingroup\@sanitize@url \@url }%
\providecommand \@url [1]{\endgroup\@href {#1}{\urlprefix }}%
\providecommand \urlprefix  [0]{URL }%
\providecommand \Eprint [0]{\href }%
\providecommand \doibase [0]{http://dx.doi.org/}%
\providecommand \selectlanguage [0]{\@gobble}%
\providecommand \bibinfo  [0]{\@secondoftwo}%
\providecommand \bibfield  [0]{\@secondoftwo}%
\providecommand \translation [1]{[#1]}%
\providecommand \BibitemOpen [0]{}%
\providecommand \bibitemStop [0]{}%
\providecommand \bibitemNoStop [0]{.\EOS\space}%
\providecommand \EOS [0]{\spacefactor3000\relax}%
\providecommand \BibitemShut  [1]{\csname bibitem#1\endcsname}%
\let\auto@bib@innerbib\@empty
\bibitem [{\citenamefont {Kharzeev}\ \emph {et~al.}(2008)\citenamefont
  {Kharzeev}, \citenamefont {McLerran},\ and\ \citenamefont
  {Warringa}}]{Kharzeev:2007jp}%
  \BibitemOpen
  \bibfield  {author} {\bibinfo {author} {\bibfnamefont {D.~E.}\ \bibnamefont
  {Kharzeev}}, \bibinfo {author} {\bibfnamefont {L.~D.}\ \bibnamefont
  {McLerran}}, \ and\ \bibinfo {author} {\bibfnamefont {H.~J.}\ \bibnamefont
  {Warringa}},\ }\href {\doibase 10.1016/j.nuclphysa.2008.02.298} {\bibfield
  {journal} {\bibinfo  {journal} {Nucl. Phys.}\ }\textbf {\bibinfo {volume}
  {A803}},\ \bibinfo {pages} {227} (\bibinfo {year} {2008})},\ \Eprint
  {http://arxiv.org/abs/0711.0950} {arXiv:0711.0950 [hep-ph]} \BibitemShut
  {NoStop}%
\bibitem [{\citenamefont {Skokov}\ \emph {et~al.}(2009)\citenamefont {Skokov},
  \citenamefont {Illarionov},\ and\ \citenamefont {Toneev}}]{Skokov:2009qp}%
  \BibitemOpen
  \bibfield  {author} {\bibinfo {author} {\bibfnamefont {V.}~\bibnamefont
  {Skokov}}, \bibinfo {author} {\bibfnamefont {A.~{\relax Yu}.}\ \bibnamefont
  {Illarionov}}, \ and\ \bibinfo {author} {\bibfnamefont {V.}~\bibnamefont
  {Toneev}},\ }\href {\doibase 10.1142/S0217751X09047570} {\bibfield  {journal}
  {\bibinfo  {journal} {Int. J. Mod. Phys.}\ }\textbf {\bibinfo {volume}
  {A24}},\ \bibinfo {pages} {5925} (\bibinfo {year} {2009})},\ \Eprint
  {http://arxiv.org/abs/0907.1396} {arXiv:0907.1396 [nucl-th]} \BibitemShut
  {NoStop}%
\bibitem [{\citenamefont {Voronyuk}\ \emph {et~al.}(2011)\citenamefont
  {Voronyuk}, \citenamefont {Toneev}, \citenamefont {Cassing}, \citenamefont
  {Bratkovskaya}, \citenamefont {Konchakovski},\ and\ \citenamefont
  {Voloshin}}]{Voronyuk:2011jd}%
  \BibitemOpen
  \bibfield  {author} {\bibinfo {author} {\bibfnamefont {V.}~\bibnamefont
  {Voronyuk}}, \bibinfo {author} {\bibfnamefont {V.~D.}\ \bibnamefont
  {Toneev}}, \bibinfo {author} {\bibfnamefont {W.}~\bibnamefont {Cassing}},
  \bibinfo {author} {\bibfnamefont {E.~L.}\ \bibnamefont {Bratkovskaya}},
  \bibinfo {author} {\bibfnamefont {V.~P.}\ \bibnamefont {Konchakovski}}, \
  and\ \bibinfo {author} {\bibfnamefont {S.~A.}\ \bibnamefont {Voloshin}},\
  }\href {\doibase 10.1103/PhysRevC.83.054911} {\bibfield  {journal} {\bibinfo
  {journal} {Phys. Rev.}\ }\textbf {\bibinfo {volume} {C83}},\ \bibinfo {pages}
  {054911} (\bibinfo {year} {2011})},\ \Eprint {http://arxiv.org/abs/1103.4239}
  {arXiv:1103.4239 [nucl-th]} \BibitemShut {NoStop}%
\bibitem [{\citenamefont {Deng}\ and\ \citenamefont
  {Huang}(2012)}]{Deng:2012pc}%
  \BibitemOpen
  \bibfield  {author} {\bibinfo {author} {\bibfnamefont {W.-T.}\ \bibnamefont
  {Deng}}\ and\ \bibinfo {author} {\bibfnamefont {X.-G.}\ \bibnamefont
  {Huang}},\ }\href {\doibase 10.1103/PhysRevC.85.044907} {\bibfield  {journal}
  {\bibinfo  {journal} {Phys. Rev.}\ }\textbf {\bibinfo {volume} {C85}},\
  \bibinfo {pages} {044907} (\bibinfo {year} {2012})},\ \Eprint
  {http://arxiv.org/abs/1201.5108} {arXiv:1201.5108 [nucl-th]} \BibitemShut
  {NoStop}%
\bibitem [{\citenamefont {Alford}\ and\ \citenamefont
  {Strickland}(2013)}]{Alford:2013jva}%
  \BibitemOpen
  \bibfield  {author} {\bibinfo {author} {\bibfnamefont {J.}~\bibnamefont
  {Alford}}\ and\ \bibinfo {author} {\bibfnamefont {M.}~\bibnamefont
  {Strickland}},\ }\href {\doibase 10.1103/PhysRevD.88.105017} {\bibfield
  {journal} {\bibinfo  {journal} {Phys.Rev.}\ }\textbf {\bibinfo {volume}
  {D88}},\ \bibinfo {pages} {105017} (\bibinfo {year} {2013})},\ \Eprint
  {http://arxiv.org/abs/1309.3003} {arXiv:1309.3003 [hep-ph]} \BibitemShut
  {NoStop}%
\bibitem [{\citenamefont {Bonati}\ \emph {et~al.}(2015)\citenamefont {Bonati},
  \citenamefont {D'Elia},\ and\ \citenamefont {Rucci}}]{Bonati:2015dka}%
  \BibitemOpen
  \bibfield  {author} {\bibinfo {author} {\bibfnamefont {C.}~\bibnamefont
  {Bonati}}, \bibinfo {author} {\bibfnamefont {M.}~\bibnamefont {D'Elia}}, \
  and\ \bibinfo {author} {\bibfnamefont {A.}~\bibnamefont {Rucci}},\ }\href
  {\doibase 10.1103/PhysRevD.92.054014} {\bibfield  {journal} {\bibinfo
  {journal} {Phys. Rev.}\ }\textbf {\bibinfo {volume} {D92}},\ \bibinfo {pages}
  {054014} (\bibinfo {year} {2015})},\ \Eprint
  {http://arxiv.org/abs/1506.07890} {arXiv:1506.07890 [hep-ph]} \BibitemShut
  {NoStop}%
\bibitem [{\citenamefont {Suzuki}\ and\ \citenamefont
  {Yoshida}(2016)}]{Suzuki:2016kcs}%
  \BibitemOpen
  \bibfield  {author} {\bibinfo {author} {\bibfnamefont {K.}~\bibnamefont
  {Suzuki}}\ and\ \bibinfo {author} {\bibfnamefont {T.}~\bibnamefont
  {Yoshida}},\ }\href {\doibase 10.1103/PhysRevD.93.051502} {\bibfield
  {journal} {\bibinfo  {journal} {Phys. Rev.}\ }\textbf {\bibinfo {volume}
  {D93}},\ \bibinfo {pages} {051502} (\bibinfo {year} {2016})},\ \Eprint
  {http://arxiv.org/abs/1601.02178} {arXiv:1601.02178 [hep-ph]} \BibitemShut
  {NoStop}%
\bibitem [{\citenamefont {Yoshida}\ and\ \citenamefont
  {Suzuki}(2016)}]{Yoshida:2016xgm}%
  \BibitemOpen
  \bibfield  {author} {\bibinfo {author} {\bibfnamefont {T.}~\bibnamefont
  {Yoshida}}\ and\ \bibinfo {author} {\bibfnamefont {K.}~\bibnamefont
  {Suzuki}},\ }\href {\doibase 10.1103/PhysRevD.94.074043} {\bibfield
  {journal} {\bibinfo  {journal} {Phys. Rev.}\ }\textbf {\bibinfo {volume}
  {D94}},\ \bibinfo {pages} {074043} (\bibinfo {year} {2016})},\ \Eprint
  {http://arxiv.org/abs/1607.04935} {arXiv:1607.04935 [hep-ph]} \BibitemShut
  {NoStop}%
\bibitem [{\citenamefont {Cho}\ \emph {et~al.}(2014)\citenamefont {Cho},
  \citenamefont {Hattori}, \citenamefont {Lee}, \citenamefont {Morita},\ and\
  \citenamefont {Ozaki}}]{Cho:2014exa}%
  \BibitemOpen
  \bibfield  {author} {\bibinfo {author} {\bibfnamefont {S.}~\bibnamefont
  {Cho}}, \bibinfo {author} {\bibfnamefont {K.}~\bibnamefont {Hattori}},
  \bibinfo {author} {\bibfnamefont {S.~H.}\ \bibnamefont {Lee}}, \bibinfo
  {author} {\bibfnamefont {K.}~\bibnamefont {Morita}}, \ and\ \bibinfo {author}
  {\bibfnamefont {S.}~\bibnamefont {Ozaki}},\ }\href {\doibase
  10.1103/PhysRevLett.113.172301} {\bibfield  {journal} {\bibinfo  {journal}
  {Phys. Rev. Lett.}\ }\textbf {\bibinfo {volume} {113}},\ \bibinfo {pages}
  {172301} (\bibinfo {year} {2014})},\ \Eprint {http://arxiv.org/abs/1406.4586}
  {arXiv:1406.4586 [hep-ph]} \BibitemShut {NoStop}%
\bibitem [{\citenamefont {Cho}\ \emph {et~al.}(2015)\citenamefont {Cho},
  \citenamefont {Hattori}, \citenamefont {Lee}, \citenamefont {Morita},\ and\
  \citenamefont {Ozaki}}]{Cho:2014loa}%
  \BibitemOpen
  \bibfield  {author} {\bibinfo {author} {\bibfnamefont {S.}~\bibnamefont
  {Cho}}, \bibinfo {author} {\bibfnamefont {K.}~\bibnamefont {Hattori}},
  \bibinfo {author} {\bibfnamefont {S.~H.}\ \bibnamefont {Lee}}, \bibinfo
  {author} {\bibfnamefont {K.}~\bibnamefont {Morita}}, \ and\ \bibinfo {author}
  {\bibfnamefont {S.}~\bibnamefont {Ozaki}},\ }\href {\doibase
  10.1103/PhysRevD.91.045025} {\bibfield  {journal} {\bibinfo  {journal} {Phys.
  Rev.}\ }\textbf {\bibinfo {volume} {D91}},\ \bibinfo {pages} {045025}
  (\bibinfo {year} {2015})},\ \Eprint {http://arxiv.org/abs/1411.7675}
  {arXiv:1411.7675 [hep-ph]} \BibitemShut {NoStop}%
\bibitem [{\citenamefont {Miransky}\ and\ \citenamefont
  {Shovkovy}(2002)}]{Miransky:2002rp}%
  \BibitemOpen
  \bibfield  {author} {\bibinfo {author} {\bibfnamefont {V.~A.}\ \bibnamefont
  {Miransky}}\ and\ \bibinfo {author} {\bibfnamefont {I.~A.}\ \bibnamefont
  {Shovkovy}},\ }\href {\doibase 10.1103/PhysRevD.66.045006} {\bibfield
  {journal} {\bibinfo  {journal} {Phys. Rev.}\ }\textbf {\bibinfo {volume}
  {D66}},\ \bibinfo {pages} {045006} (\bibinfo {year} {2002})},\ \Eprint
  {http://arxiv.org/abs/hep-ph/0205348} {arXiv:hep-ph/0205348 [hep-ph]}
  \BibitemShut {NoStop}%
\bibitem [{\citenamefont {Andreichikov}\ \emph {et~al.}(2013)\citenamefont
  {Andreichikov}, \citenamefont {Orlovsky},\ and\ \citenamefont
  {Simonov}}]{Andreichikov:2012xe}%
  \BibitemOpen
  \bibfield  {author} {\bibinfo {author} {\bibfnamefont {M.~A.}\ \bibnamefont
  {Andreichikov}}, \bibinfo {author} {\bibfnamefont {V.~D.}\ \bibnamefont
  {Orlovsky}}, \ and\ \bibinfo {author} {\bibfnamefont {{\relax Yu}.~A.}\
  \bibnamefont {Simonov}},\ }\href {\doibase 10.1103/PhysRevLett.110.162002}
  {\bibfield  {journal} {\bibinfo  {journal} {Phys. Rev. Lett.}\ }\textbf
  {\bibinfo {volume} {110}},\ \bibinfo {pages} {162002} (\bibinfo {year}
  {2013})},\ \Eprint {http://arxiv.org/abs/1211.6568} {arXiv:1211.6568
  [hep-ph]} \BibitemShut {NoStop}%
\bibitem [{\citenamefont {Chernodub}(2014)}]{Chernodub:2014uua}%
  \BibitemOpen
  \bibfield  {author} {\bibinfo {author} {\bibfnamefont {M.~N.}\ \bibnamefont
  {Chernodub}},\ }\href {\doibase 10.1142/S0217732314501624} {\bibfield
  {journal} {\bibinfo  {journal} {Mod. Phys. Lett.}\ }\textbf {\bibinfo
  {volume} {A29}},\ \bibinfo {pages} {1450162} (\bibinfo {year}
  {2014})}\BibitemShut {NoStop}%
\bibitem [{\citenamefont {Bonati}\ \emph {et~al.}(2014)\citenamefont {Bonati},
  \citenamefont {D'Elia}, \citenamefont {Mariti}, \citenamefont {Mesiti},
  \citenamefont {Negro},\ and\ \citenamefont {Sanfilippo}}]{Bonati:2014ksa}%
  \BibitemOpen
  \bibfield  {author} {\bibinfo {author} {\bibfnamefont {C.}~\bibnamefont
  {Bonati}}, \bibinfo {author} {\bibfnamefont {M.}~\bibnamefont {D'Elia}},
  \bibinfo {author} {\bibfnamefont {M.}~\bibnamefont {Mariti}}, \bibinfo
  {author} {\bibfnamefont {M.}~\bibnamefont {Mesiti}}, \bibinfo {author}
  {\bibfnamefont {F.}~\bibnamefont {Negro}}, \ and\ \bibinfo {author}
  {\bibfnamefont {F.}~\bibnamefont {Sanfilippo}},\ }\href {\doibase
  10.1103/PhysRevD.89.114502} {\bibfield  {journal} {\bibinfo  {journal} {Phys.
  Rev.}\ }\textbf {\bibinfo {volume} {D89}},\ \bibinfo {pages} {114502}
  (\bibinfo {year} {2014})},\ \Eprint {http://arxiv.org/abs/1403.6094}
  {arXiv:1403.6094 [hep-lat]} \BibitemShut {NoStop}%
\bibitem [{\citenamefont {Rougemont}\ \emph {et~al.}(2015)\citenamefont
  {Rougemont}, \citenamefont {Critelli},\ and\ \citenamefont
  {Noronha}}]{Rougemont:2014efa}%
  \BibitemOpen
  \bibfield  {author} {\bibinfo {author} {\bibfnamefont {R.}~\bibnamefont
  {Rougemont}}, \bibinfo {author} {\bibfnamefont {R.}~\bibnamefont {Critelli}},
  \ and\ \bibinfo {author} {\bibfnamefont {J.}~\bibnamefont {Noronha}},\ }\href
  {\doibase 10.1103/PhysRevD.91.066001} {\bibfield  {journal} {\bibinfo
  {journal} {Phys. Rev.}\ }\textbf {\bibinfo {volume} {D91}},\ \bibinfo {pages}
  {066001} (\bibinfo {year} {2015})},\ \Eprint {http://arxiv.org/abs/1409.0556}
  {arXiv:1409.0556 [hep-th]} \BibitemShut {NoStop}%
\bibitem [{\citenamefont {Simonov}\ and\ \citenamefont
  {Trusov}(2015)}]{Simonov:2015yka}%
  \BibitemOpen
  \bibfield  {author} {\bibinfo {author} {\bibfnamefont {{\relax Yu}.~A.}\
  \bibnamefont {Simonov}}\ and\ \bibinfo {author} {\bibfnamefont {M.~A.}\
  \bibnamefont {Trusov}},\ }\href {\doibase 10.1016/j.physletb.2015.05.032}
  {\bibfield  {journal} {\bibinfo  {journal} {Phys. Lett.}\ }\textbf {\bibinfo
  {volume} {B747}},\ \bibinfo {pages} {48} (\bibinfo {year} {2015})},\ \Eprint
  {http://arxiv.org/abs/1503.08531} {arXiv:1503.08531 [hep-ph]} \BibitemShut
  {NoStop}%
\bibitem [{\citenamefont {Bonati}\ \emph {et~al.}(2016)\citenamefont {Bonati},
  \citenamefont {D'Elia}, \citenamefont {Mariti}, \citenamefont {Mesiti},
  \citenamefont {Negro}, \citenamefont {Rucci},\ and\ \citenamefont
  {Sanfilippo}}]{Bonati:2016kxj}%
  \BibitemOpen
  \bibfield  {author} {\bibinfo {author} {\bibfnamefont {C.}~\bibnamefont
  {Bonati}}, \bibinfo {author} {\bibfnamefont {M.}~\bibnamefont {D'Elia}},
  \bibinfo {author} {\bibfnamefont {M.}~\bibnamefont {Mariti}}, \bibinfo
  {author} {\bibfnamefont {M.}~\bibnamefont {Mesiti}}, \bibinfo {author}
  {\bibfnamefont {F.}~\bibnamefont {Negro}}, \bibinfo {author} {\bibfnamefont
  {A.}~\bibnamefont {Rucci}}, \ and\ \bibinfo {author} {\bibfnamefont
  {F.}~\bibnamefont {Sanfilippo}},\ }\href {\doibase
  10.1103/PhysRevD.94.094007} {\bibfield  {journal} {\bibinfo  {journal} {Phys.
  Rev.}\ }\textbf {\bibinfo {volume} {D94}},\ \bibinfo {pages} {094007}
  (\bibinfo {year} {2016})},\ \Eprint {http://arxiv.org/abs/1607.08160}
  {arXiv:1607.08160 [hep-lat]} \BibitemShut {NoStop}%
\bibitem [{\citenamefont {Fukushima}\ \emph {et~al.}(2016)\citenamefont
  {Fukushima}, \citenamefont {Hattori}, \citenamefont {Yee},\ and\
  \citenamefont {Yin}}]{Fukushima:2015wck}%
  \BibitemOpen
  \bibfield  {author} {\bibinfo {author} {\bibfnamefont {K.}~\bibnamefont
  {Fukushima}}, \bibinfo {author} {\bibfnamefont {K.}~\bibnamefont {Hattori}},
  \bibinfo {author} {\bibfnamefont {H.-U.}\ \bibnamefont {Yee}}, \ and\
  \bibinfo {author} {\bibfnamefont {Y.}~\bibnamefont {Yin}},\ }\href {\doibase
  10.1103/PhysRevD.93.074028} {\bibfield  {journal} {\bibinfo  {journal} {Phys.
  Rev.}\ }\textbf {\bibinfo {volume} {D93}},\ \bibinfo {pages} {074028}
  (\bibinfo {year} {2016})},\ \Eprint {http://arxiv.org/abs/1512.03689}
  {arXiv:1512.03689 [hep-ph]} \BibitemShut {NoStop}%
\bibitem [{\citenamefont {Finazzo}\ \emph {et~al.}(2016)\citenamefont
  {Finazzo}, \citenamefont {Critelli}, \citenamefont {Rougemont},\ and\
  \citenamefont {Noronha}}]{Finazzo:2016mhm}%
  \BibitemOpen
  \bibfield  {author} {\bibinfo {author} {\bibfnamefont {S.~I.}\ \bibnamefont
  {Finazzo}}, \bibinfo {author} {\bibfnamefont {R.}~\bibnamefont {Critelli}},
  \bibinfo {author} {\bibfnamefont {R.}~\bibnamefont {Rougemont}}, \ and\
  \bibinfo {author} {\bibfnamefont {J.}~\bibnamefont {Noronha}},\ }\href
  {\doibase 10.1103/PhysRevD.94.054020} {\bibfield  {journal} {\bibinfo
  {journal} {Phys. Rev.}\ }\textbf {\bibinfo {volume} {D94}},\ \bibinfo {pages}
  {054020} (\bibinfo {year} {2016})},\ \Eprint
  {http://arxiv.org/abs/1605.06061} {arXiv:1605.06061 [hep-ph]} \BibitemShut
  {NoStop}%
\bibitem [{\citenamefont {Das}\ \emph {et~al.}(2017)\citenamefont {Das},
  \citenamefont {Plumari}, \citenamefont {Chatterjee}, \citenamefont {Alam},
  \citenamefont {Scardina},\ and\ \citenamefont {Greco}}]{Das:2016cwd}%
  \BibitemOpen
  \bibfield  {author} {\bibinfo {author} {\bibfnamefont {S.~K.}\ \bibnamefont
  {Das}}, \bibinfo {author} {\bibfnamefont {S.}~\bibnamefont {Plumari}},
  \bibinfo {author} {\bibfnamefont {S.}~\bibnamefont {Chatterjee}}, \bibinfo
  {author} {\bibfnamefont {J.}~\bibnamefont {Alam}}, \bibinfo {author}
  {\bibfnamefont {F.}~\bibnamefont {Scardina}}, \ and\ \bibinfo {author}
  {\bibfnamefont {V.}~\bibnamefont {Greco}},\ }\href {\doibase
  10.1016/j.physletb.2017.02.046} {\bibfield  {journal} {\bibinfo  {journal}
  {Phys. Lett.}\ }\textbf {\bibinfo {volume} {B768}},\ \bibinfo {pages} {260}
  (\bibinfo {year} {2017})},\ \Eprint {http://arxiv.org/abs/1608.02231}
  {arXiv:1608.02231 [nucl-th]} \BibitemShut {NoStop}%
\bibitem [{\citenamefont {Machado}\ \emph {et~al.}(2013)\citenamefont
  {Machado}, \citenamefont {Navarra}, \citenamefont {de~Oliveira},
  \citenamefont {Noronha},\ and\ \citenamefont {Strickland}}]{Machado:2013rta}%
  \BibitemOpen
  \bibfield  {author} {\bibinfo {author} {\bibfnamefont {C.~S.}\ \bibnamefont
  {Machado}}, \bibinfo {author} {\bibfnamefont {F.~S.}\ \bibnamefont
  {Navarra}}, \bibinfo {author} {\bibfnamefont {E.~G.}\ \bibnamefont
  {de~Oliveira}}, \bibinfo {author} {\bibfnamefont {J.}~\bibnamefont
  {Noronha}}, \ and\ \bibinfo {author} {\bibfnamefont {M.}~\bibnamefont
  {Strickland}},\ }\href {\doibase 10.1103/PhysRevD.88.034009} {\bibfield
  {journal} {\bibinfo  {journal} {Phys. Rev.}\ }\textbf {\bibinfo {volume}
  {D88}},\ \bibinfo {pages} {034009} (\bibinfo {year} {2013})},\ \Eprint
  {http://arxiv.org/abs/1305.3308} {arXiv:1305.3308 [hep-ph]} \BibitemShut
  {NoStop}%
\bibitem [{\citenamefont {Machado}\ \emph {et~al.}(2014)\citenamefont
  {Machado}, \citenamefont {Matheus}, \citenamefont {Finazzo},\ and\
  \citenamefont {Noronha}}]{Machado:2013yaa}%
  \BibitemOpen
  \bibfield  {author} {\bibinfo {author} {\bibfnamefont {C.~S.}\ \bibnamefont
  {Machado}}, \bibinfo {author} {\bibfnamefont {R.~D.}\ \bibnamefont
  {Matheus}}, \bibinfo {author} {\bibfnamefont {S.~I.}\ \bibnamefont
  {Finazzo}}, \ and\ \bibinfo {author} {\bibfnamefont {J.}~\bibnamefont
  {Noronha}},\ }\href {\doibase 10.1103/PhysRevD.89.074027} {\bibfield
  {journal} {\bibinfo  {journal} {Phys. Rev.}\ }\textbf {\bibinfo {volume}
  {D89}},\ \bibinfo {pages} {074027} (\bibinfo {year} {2014})},\ \Eprint
  {http://arxiv.org/abs/1307.1797} {arXiv:1307.1797 [hep-ph]} \BibitemShut
  {NoStop}%
\bibitem [{\citenamefont {Gubler}\ \emph {et~al.}(2016)\citenamefont {Gubler},
  \citenamefont {Hattori}, \citenamefont {Lee}, \citenamefont {Oka},
  \citenamefont {Ozaki},\ and\ \citenamefont {Suzuki}}]{Gubler:2015qok}%
  \BibitemOpen
  \bibfield  {author} {\bibinfo {author} {\bibfnamefont {P.}~\bibnamefont
  {Gubler}}, \bibinfo {author} {\bibfnamefont {K.}~\bibnamefont {Hattori}},
  \bibinfo {author} {\bibfnamefont {S.~H.}\ \bibnamefont {Lee}}, \bibinfo
  {author} {\bibfnamefont {M.}~\bibnamefont {Oka}}, \bibinfo {author}
  {\bibfnamefont {S.}~\bibnamefont {Ozaki}}, \ and\ \bibinfo {author}
  {\bibfnamefont {K.}~\bibnamefont {Suzuki}},\ }\href {\doibase
  10.1103/PhysRevD.93.054026} {\bibfield  {journal} {\bibinfo  {journal} {Phys.
  Rev.}\ }\textbf {\bibinfo {volume} {D93}},\ \bibinfo {pages} {054026}
  (\bibinfo {year} {2016})},\ \Eprint {http://arxiv.org/abs/1512.08864}
  {arXiv:1512.08864 [hep-ph]} \BibitemShut {NoStop}%
\bibitem [{\citenamefont {Hattori}\ and\ \citenamefont
  {Huang}(2017)}]{Hattori:2016emy}%
  \BibitemOpen
  \bibfield  {author} {\bibinfo {author} {\bibfnamefont {K.}~\bibnamefont
  {Hattori}}\ and\ \bibinfo {author} {\bibfnamefont {X.-G.}\ \bibnamefont
  {Huang}},\ }\href {\doibase 10.1007/s41365-016-0178-3} {\bibfield  {journal}
  {\bibinfo  {journal} {Nucl. Sci. Tech.}\ }\textbf {\bibinfo {volume} {28}},\
  \bibinfo {pages} {26} (\bibinfo {year} {2017})},\ \Eprint
  {http://arxiv.org/abs/1609.00747} {arXiv:1609.00747 [nucl-th]} \BibitemShut
  {NoStop}%
\bibitem [{\citenamefont {Karsch}\ and\ \citenamefont
  {Petronzio}(1988)}]{Karsch:1987zw}%
  \BibitemOpen
  \bibfield  {author} {\bibinfo {author} {\bibfnamefont {F.}~\bibnamefont
  {Karsch}}\ and\ \bibinfo {author} {\bibfnamefont {R.}~\bibnamefont
  {Petronzio}},\ }\href {\doibase 10.1007/BF01549724} {\bibfield  {journal}
  {\bibinfo  {journal} {Z. Phys.}\ }\textbf {\bibinfo {volume} {C37}},\
  \bibinfo {pages} {627} (\bibinfo {year} {1988})}\BibitemShut {NoStop}%
\bibitem [{\citenamefont {Blaizot}\ and\ \citenamefont
  {Ollitrault}(1989)}]{Blaizot:1988ec}%
  \BibitemOpen
  \bibfield  {author} {\bibinfo {author} {\bibfnamefont {J.~P.}\ \bibnamefont
  {Blaizot}}\ and\ \bibinfo {author} {\bibfnamefont {J.-Y.}\ \bibnamefont
  {Ollitrault}},\ }\href {\doibase 10.1103/PhysRevD.39.232} {\bibfield
  {journal} {\bibinfo  {journal} {Phys. Rev.}\ }\textbf {\bibinfo {volume}
  {D39}},\ \bibinfo {pages} {232} (\bibinfo {year} {1989})}\BibitemShut
  {NoStop}%
\bibitem [{\citenamefont {Kharzeev}\ and\ \citenamefont
  {Thews}(1999)}]{Kharzeev:1999bh}%
  \BibitemOpen
  \bibfield  {author} {\bibinfo {author} {\bibfnamefont {D.}~\bibnamefont
  {Kharzeev}}\ and\ \bibinfo {author} {\bibfnamefont {R.~L.}\ \bibnamefont
  {Thews}},\ }\href {\doibase 10.1103/PhysRevC.60.041901} {\bibfield  {journal}
  {\bibinfo  {journal} {Phys. Rev.}\ }\textbf {\bibinfo {volume} {C60}},\
  \bibinfo {pages} {041901} (\bibinfo {year} {1999})},\ \Eprint
  {http://arxiv.org/abs/nucl-th/9907021} {arXiv:nucl-th/9907021 [nucl-th]}
  \BibitemShut {NoStop}%
\bibitem [{\citenamefont {Song}\ \emph {et~al.}(2013)\citenamefont {Song},
  \citenamefont {Ko},\ and\ \citenamefont {Lee}}]{Song:2013lov}%
  \BibitemOpen
  \bibfield  {author} {\bibinfo {author} {\bibfnamefont {T.}~\bibnamefont
  {Song}}, \bibinfo {author} {\bibfnamefont {C.~M.}\ \bibnamefont {Ko}}, \ and\
  \bibinfo {author} {\bibfnamefont {S.~H.}\ \bibnamefont {Lee}},\ }\href
  {\doibase 10.1103/PhysRevC.87.034910} {\bibfield  {journal} {\bibinfo
  {journal} {Phys. Rev.}\ }\textbf {\bibinfo {volume} {C87}},\ \bibinfo {pages}
  {034910} (\bibinfo {year} {2013})},\ \Eprint {http://arxiv.org/abs/1302.4395}
  {arXiv:1302.4395 [nucl-th]} \BibitemShut {NoStop}%
\bibitem [{\citenamefont {Song}\ \emph {et~al.}(2015)\citenamefont {Song},
  \citenamefont {Ko},\ and\ \citenamefont {Lee}}]{Song:2015bja}%
  \BibitemOpen
  \bibfield  {author} {\bibinfo {author} {\bibfnamefont {T.}~\bibnamefont
  {Song}}, \bibinfo {author} {\bibfnamefont {C.~M.}\ \bibnamefont {Ko}}, \ and\
  \bibinfo {author} {\bibfnamefont {S.~H.}\ \bibnamefont {Lee}},\ }\href
  {\doibase 10.1103/PhysRevC.91.044909} {\bibfield  {journal} {\bibinfo
  {journal} {Phys. Rev.}\ }\textbf {\bibinfo {volume} {C91}},\ \bibinfo {pages}
  {044909} (\bibinfo {year} {2015})},\ \Eprint
  {http://arxiv.org/abs/1502.05734} {arXiv:1502.05734 [nucl-th]} \BibitemShut
  {NoStop}%
\bibitem [{\citenamefont {Shuryak}(1984)}]{Shuryak:1983jg}%
  \BibitemOpen
  \bibfield  {author} {\bibinfo {author} {\bibfnamefont {E.~V.}\ \bibnamefont
  {Shuryak}},\ }\href {\doibase 10.1016/0370-2693(84)91159-6} {\bibfield
  {journal} {\bibinfo  {journal} {Phys. Lett.}\ }\textbf {\bibinfo {volume}
  {B136}},\ \bibinfo {pages} {269} (\bibinfo {year} {1984})}\BibitemShut
  {NoStop}%
\bibitem [{\citenamefont {Huang}(2016)}]{Huang:2015oca}%
  \BibitemOpen
  \bibfield  {author} {\bibinfo {author} {\bibfnamefont {X.-G.}\ \bibnamefont
  {Huang}},\ }\href {\doibase 10.1088/0034-4885/79/7/076302} {\bibfield
  {journal} {\bibinfo  {journal} {Rept. Prog. Phys.}\ }\textbf {\bibinfo
  {volume} {79}},\ \bibinfo {pages} {076302} (\bibinfo {year} {2016})},\
  \Eprint {http://arxiv.org/abs/1509.04073} {arXiv:1509.04073 [nucl-th]}
  \BibitemShut {NoStop}%
\bibitem [{\citenamefont {Tuchin}(2013{\natexlab{a}})}]{Tuchin:2013ie}%
  \BibitemOpen
  \bibfield  {author} {\bibinfo {author} {\bibfnamefont {K.}~\bibnamefont
  {Tuchin}},\ }\href {\doibase 10.1155/2013/490495} {\bibfield  {journal}
  {\bibinfo  {journal} {Adv. High Energy Phys.}\ }\textbf {\bibinfo {volume}
  {2013}},\ \bibinfo {pages} {490495} (\bibinfo {year} {2013}{\natexlab{a}})},\
  \Eprint {http://arxiv.org/abs/1301.0099} {arXiv:1301.0099 [hep-ph]}
  \BibitemShut {NoStop}%
\bibitem [{\citenamefont {McLerran}\ and\ \citenamefont
  {Skokov}(2014)}]{McLerran:2013hla}%
  \BibitemOpen
  \bibfield  {author} {\bibinfo {author} {\bibfnamefont {L.}~\bibnamefont
  {McLerran}}\ and\ \bibinfo {author} {\bibfnamefont {V.}~\bibnamefont
  {Skokov}},\ }\href {\doibase 10.1016/j.nuclphysa.2014.05.008} {\bibfield
  {journal} {\bibinfo  {journal} {Nucl. Phys.}\ }\textbf {\bibinfo {volume}
  {A929}},\ \bibinfo {pages} {184} (\bibinfo {year} {2014})},\ \Eprint
  {http://arxiv.org/abs/1305.0774} {arXiv:1305.0774 [hep-ph]} \BibitemShut
  {NoStop}%
\bibitem [{\citenamefont {Tuchin}(2013{\natexlab{b}})}]{Tuchin:2013apa}%
  \BibitemOpen
  \bibfield  {author} {\bibinfo {author} {\bibfnamefont {K.}~\bibnamefont
  {Tuchin}},\ }\href {\doibase 10.1103/PhysRevC.88.024911} {\bibfield
  {journal} {\bibinfo  {journal} {Phys. Rev.}\ }\textbf {\bibinfo {volume}
  {C88}},\ \bibinfo {pages} {024911} (\bibinfo {year} {2013}{\natexlab{b}})},\
  \Eprint {http://arxiv.org/abs/1305.5806} {arXiv:1305.5806 [hep-ph]}
  \BibitemShut {NoStop}%
\bibitem [{\citenamefont {G{\" u}rsoy}\ \emph {et~al.}(2014)\citenamefont {G{\"
  u}rsoy}, \citenamefont {Kharzeev},\ and\ \citenamefont
  {Rajagopal}}]{Gursoy:2014aka}%
  \BibitemOpen
  \bibfield  {author} {\bibinfo {author} {\bibfnamefont {U.}~\bibnamefont {G{\"
  u}rsoy}}, \bibinfo {author} {\bibfnamefont {D.}~\bibnamefont {Kharzeev}}, \
  and\ \bibinfo {author} {\bibfnamefont {K.}~\bibnamefont {Rajagopal}},\ }\href
  {\doibase 10.1103/PhysRevC.89.054905} {\bibfield  {journal} {\bibinfo
  {journal} {Phys. Rev.}\ }\textbf {\bibinfo {volume} {C89}},\ \bibinfo {pages}
  {054905} (\bibinfo {year} {2014})},\ \Eprint {http://arxiv.org/abs/1401.3805}
  {arXiv:1401.3805 [hep-ph]} \BibitemShut {NoStop}%
\bibitem [{\citenamefont {Zakharov}(2014)}]{Zakharov:2014dia}%
  \BibitemOpen
  \bibfield  {author} {\bibinfo {author} {\bibfnamefont {B.~G.}\ \bibnamefont
  {Zakharov}},\ }\href {\doibase 10.1016/j.physletb.2014.08.068} {\bibfield
  {journal} {\bibinfo  {journal} {Phys. Lett.}\ }\textbf {\bibinfo {volume}
  {B737}},\ \bibinfo {pages} {262} (\bibinfo {year} {2014})},\ \Eprint
  {http://arxiv.org/abs/1404.5047} {arXiv:1404.5047 [hep-ph]} \BibitemShut
  {NoStop}%
\bibitem [{\citenamefont {Tuchin}(2016)}]{Tuchin:2015oka}%
  \BibitemOpen
  \bibfield  {author} {\bibinfo {author} {\bibfnamefont {K.}~\bibnamefont
  {Tuchin}},\ }\href {\doibase 10.1103/PhysRevC.93.014905} {\bibfield
  {journal} {\bibinfo  {journal} {Phys. Rev.}\ }\textbf {\bibinfo {volume}
  {C93}},\ \bibinfo {pages} {014905} (\bibinfo {year} {2016})},\ \Eprint
  {http://arxiv.org/abs/1508.06925} {arXiv:1508.06925 [hep-ph]} \BibitemShut
  {NoStop}%
\bibitem [{\citenamefont {Buividovich}\ \emph {et~al.}(2010)\citenamefont
  {Buividovich}, \citenamefont {Chernodub}, \citenamefont {Kharzeev},
  \citenamefont {Kalaydzhyan}, \citenamefont {Luschevskaya},\ and\
  \citenamefont {Polikarpov}}]{Buividovich:2010tn}%
  \BibitemOpen
  \bibfield  {author} {\bibinfo {author} {\bibfnamefont {P.~V.}\ \bibnamefont
  {Buividovich}}, \bibinfo {author} {\bibfnamefont {M.~N.}\ \bibnamefont
  {Chernodub}}, \bibinfo {author} {\bibfnamefont {D.~E.}\ \bibnamefont
  {Kharzeev}}, \bibinfo {author} {\bibfnamefont {T.}~\bibnamefont
  {Kalaydzhyan}}, \bibinfo {author} {\bibfnamefont {E.~V.}\ \bibnamefont
  {Luschevskaya}}, \ and\ \bibinfo {author} {\bibfnamefont {M.~I.}\
  \bibnamefont {Polikarpov}},\ }\href {\doibase 10.1103/PhysRevLett.105.132001}
  {\bibfield  {journal} {\bibinfo  {journal} {Phys. Rev. Lett.}\ }\textbf
  {\bibinfo {volume} {105}},\ \bibinfo {pages} {132001} (\bibinfo {year}
  {2010})},\ \Eprint {http://arxiv.org/abs/1003.2180} {arXiv:1003.2180
  [hep-lat]} \BibitemShut {NoStop}%
\bibitem [{\citenamefont {Hattori}\ and\ \citenamefont
  {Satow}(2016)}]{Hattori:2016cnt}%
  \BibitemOpen
  \bibfield  {author} {\bibinfo {author} {\bibfnamefont {K.}~\bibnamefont
  {Hattori}}\ and\ \bibinfo {author} {\bibfnamefont {D.}~\bibnamefont
  {Satow}},\ }\href {\doibase 10.1103/PhysRevD.94.114032} {\bibfield  {journal}
  {\bibinfo  {journal} {Phys. Rev.}\ }\textbf {\bibinfo {volume} {D94}},\
  \bibinfo {pages} {114032} (\bibinfo {year} {2016})},\ \Eprint
  {http://arxiv.org/abs/1610.06818} {arXiv:1610.06818 [hep-ph]} \BibitemShut
  {NoStop}%
\bibitem [{\citenamefont {Hattori}\ \emph {et~al.}(2017)\citenamefont
  {Hattori}, \citenamefont {Li}, \citenamefont {Satow},\ and\ \citenamefont
  {Yee}}]{Hattori:2016lqx}%
  \BibitemOpen
  \bibfield  {author} {\bibinfo {author} {\bibfnamefont {K.}~\bibnamefont
  {Hattori}}, \bibinfo {author} {\bibfnamefont {S.}~\bibnamefont {Li}},
  \bibinfo {author} {\bibfnamefont {D.}~\bibnamefont {Satow}}, \ and\ \bibinfo
  {author} {\bibfnamefont {H.-U.}\ \bibnamefont {Yee}},\ }\href {\doibase
  10.1103/PhysRevD.95.076008} {\bibfield  {journal} {\bibinfo  {journal} {Phys.
  Rev.}\ }\textbf {\bibinfo {volume} {D95}},\ \bibinfo {pages} {076008}
  (\bibinfo {year} {2017})},\ \Eprint {http://arxiv.org/abs/1610.06839}
  {arXiv:1610.06839 [hep-ph]} \BibitemShut {NoStop}%
\bibitem [{\citenamefont {Yang}\ and\ \citenamefont {M{\"
  u}ller}(2012)}]{Yang:2011cz}%
  \BibitemOpen
  \bibfield  {author} {\bibinfo {author} {\bibfnamefont {D.-L.}\ \bibnamefont
  {Yang}}\ and\ \bibinfo {author} {\bibfnamefont {B.}~\bibnamefont {M{\"
  u}ller}},\ }\href {\doibase 10.1088/0954-3899/39/1/015007} {\bibfield
  {journal} {\bibinfo  {journal} {J. Phys.}\ }\textbf {\bibinfo {volume}
  {G39}},\ \bibinfo {pages} {015007} (\bibinfo {year} {2012})},\ \Eprint
  {http://arxiv.org/abs/1108.2525} {arXiv:1108.2525 [hep-ph]} \BibitemShut
  {NoStop}%
\bibitem [{\citenamefont {Guo}\ \emph {et~al.}(2015)\citenamefont {Guo},
  \citenamefont {Shi}, \citenamefont {Xu}, \citenamefont {Xu},\ and\
  \citenamefont {Zhuang}}]{Guo:2015nsa}%
  \BibitemOpen
  \bibfield  {author} {\bibinfo {author} {\bibfnamefont {X.}~\bibnamefont
  {Guo}}, \bibinfo {author} {\bibfnamefont {S.}~\bibnamefont {Shi}}, \bibinfo
  {author} {\bibfnamefont {N.}~\bibnamefont {Xu}}, \bibinfo {author}
  {\bibfnamefont {Z.}~\bibnamefont {Xu}}, \ and\ \bibinfo {author}
  {\bibfnamefont {P.}~\bibnamefont {Zhuang}},\ }\href {\doibase
  10.1016/j.physletb.2015.10.038} {\bibfield  {journal} {\bibinfo  {journal}
  {Phys. Lett.}\ }\textbf {\bibinfo {volume} {B751}},\ \bibinfo {pages} {215}
  (\bibinfo {year} {2015})},\ \Eprint {http://arxiv.org/abs/1502.04407}
  {arXiv:1502.04407 [hep-ph]} \BibitemShut {NoStop}%
\bibitem [{\citenamefont {Dudal}\ and\ \citenamefont
  {Mertens}(2015)}]{Dudal:2014jfa}%
  \BibitemOpen
  \bibfield  {author} {\bibinfo {author} {\bibfnamefont {D.}~\bibnamefont
  {Dudal}}\ and\ \bibinfo {author} {\bibfnamefont {T.~G.}\ \bibnamefont
  {Mertens}},\ }\href {\doibase 10.1103/PhysRevD.91.086002} {\bibfield
  {journal} {\bibinfo  {journal} {Phys. Rev.}\ }\textbf {\bibinfo {volume}
  {D91}},\ \bibinfo {pages} {086002} (\bibinfo {year} {2015})},\ \Eprint
  {http://arxiv.org/abs/1410.3297} {arXiv:1410.3297 [hep-th]} \BibitemShut
  {NoStop}%
\bibitem [{\citenamefont {Marasinghe}\ and\ \citenamefont
  {Tuchin}(2011)}]{Marasinghe:2011bt}%
  \BibitemOpen
  \bibfield  {author} {\bibinfo {author} {\bibfnamefont {K.}~\bibnamefont
  {Marasinghe}}\ and\ \bibinfo {author} {\bibfnamefont {K.}~\bibnamefont
  {Tuchin}},\ }\href {\doibase 10.1103/PhysRevC.84.044908} {\bibfield
  {journal} {\bibinfo  {journal} {Phys. Rev.}\ }\textbf {\bibinfo {volume}
  {C84}},\ \bibinfo {pages} {044908} (\bibinfo {year} {2011})},\ \Eprint
  {http://arxiv.org/abs/1103.1329} {arXiv:1103.1329 [hep-ph]} \BibitemShut
  {NoStop}%
\bibitem [{\citenamefont {Sadofyev}\ and\ \citenamefont
  {Yin}(2016)}]{Sadofyev:2015hxa}%
  \BibitemOpen
  \bibfield  {author} {\bibinfo {author} {\bibfnamefont {A.~V.}\ \bibnamefont
  {Sadofyev}}\ and\ \bibinfo {author} {\bibfnamefont {Y.}~\bibnamefont {Yin}},\
  }\href {\doibase 10.1007/JHEP01(2016)052} {\bibfield  {journal} {\bibinfo
  {journal} {JHEP}\ }\textbf {\bibinfo {volume} {01}},\ \bibinfo {pages} {052}
  (\bibinfo {year} {2016})},\ \Eprint {http://arxiv.org/abs/1510.06760}
  {arXiv:1510.06760 [hep-th]} \BibitemShut {NoStop}%
\end{thebibliography}%

\end{document}